\documentclass[12pt]{iopart}

\usepackage[pdftex]{graphicx}
\usepackage{dcolumn}
\usepackage{bm}

\usepackage[T1]{fontenc}
\usepackage[latin1]{inputenc}
\usepackage[english]{babel}
\usepackage{amssymb}
\usepackage{amstext}
\usepackage{relsize}
\usepackage{subfigure}
\usepackage{color}
\usepackage{ulem}

\begin{document}

\title[Long-range quantum spin models: A QMC study.]{Quantum spin models with long-range interactions and tunnelings: A quantum Monte Carlo study.}

\author{Micha\l{} Maik$^{1,2}$, Philipp Hauke$^2$, Omjyoti Dutta$^2$, Jakub Zakrzewski$^{1,3}$ and Maciej Lewenstein$^{2,4}$}

\address{$^1$ Instytut Fizyki imienia Mariana Smoluchowskiego,
Uniwersytet Jagiello\'nski, ulica Reymonta 4, PL-30-059 Krak\'ow, Poland}

\address{$^2$ ICFO -- Institut de Ci\`encies Fot\`oniques,
Mediterranean Technology Park, E-08860 Castelldefels (Barcelona), Spain}

\address{$^3$ Mark Kac Complex Systems Research Center,
Uniwersytet Jagiello\'nski, Krak\'ow, Poland}

\address{$^4$ ICREA -- Instituci\'{o} Catalana de Recerca i Estudis Avan\c{c}ats, E-08010 Barcelona, Spain}

\ead{michal.maik@uj.edu.pl}

\date{\today}

\begin{abstract}

We use a quantum Monte Carlo method to investigate various
classes of 2D spin models with long-range interactions at low
temperatures. In particular, we study a dipolar XXZ model with
$U(1)$ symmetry that appears as a hard-core boson limit of an extended
Hubbard model describing polarized dipolar atoms or molecules in
an optical lattice. Tunneling, in such a model, is short-range,
whereas density-density couplings decay with distance following a cubic power law. We investigate also an XXZ model with long-range
couplings of all three spin components - such a model describes a
system of ultracold ions in a lattice of microtraps. We describe
an approximate phase diagram for such systems at zero and at finite
temperature, and compare their properties. In particular, we compare the extent of crystalline, superfluid, and supersolid phases. 
Our predictions apply directly to current experiments with mesoscopic numbers of polar molecules and trapped ions.

\end{abstract}


%
\maketitle


\section{Introduction}

Quantum simulators, as first proposed by Feynman \cite{feynman82}, which are
devices built to evolve according to a postulated quantum
Hamiltonian and thus ``compute'' its properties, are one of the
hot ideas which may provide a breakthrough in many-body physics.
While one must be aware of possible difficulties (see, e.g.,
\cite{hauke11a}), impressive progress has been achieved in recent
years in different systems employing cold atoms and molecules,
nuclear magnetic resonance, superconducting qubits, and ions. The
latter are extremely well controlled and already it has been
demonstrated that, indeed, quantum spin systems may be simulated
with cold-ion setups \cite{friedenauer08,kim10}. Quantum spin
systems on a lattice constitute some of the most relevant cases
for quantum simulations as there are many instances where standard
numerical techniques to compute their dynamics or even their
static behavior fail, especially for two- or three-dimensional
systems.

Chronologically the first proposition to use trapped ions to
simulate lattice spin models came from Porras and Cirac
\cite{porras04} who derived the effective spin-Hamiltonian for the
system \footnote{Although, in a different context, similar spin models were
derived earlier by Mintert and Wunderlich \cite{mintert01}.},
\begin{eqnarray}
  H = J\sum\limits_{i,j}\frac{1}{|i-j|^{3}}[\cos{\theta}(S_{i}^{z}S_{j}^{z})+\sin{\theta}(S_{i}^{x}S_{j}^{x}+S_{i}^{y}S_{j}^{y})]-\mu\sum\limits_{i} S_{i}^{z},
  \label{hamspin}
\end{eqnarray}
\noindent where $\mu$ is the chemical potential (which in this case acts as an external magnetic field), $J>0$ is
the interaction strength, and $S_{i}^{\alpha}$ are the spin operators at site $i$. All the long-ranged
interactions fall off with a $1/r^{3}$ dipolar decay, which for the ions is due to the fact that they are both induced by the same
mechanism, namely lattice vibrations mediated by the Coulomb force \cite{porras06}.
Hamiltonian \eref{hamspin} is a dipolar XXZ model where the ratio of hopping to dipolar repulsion can be scaled by $\theta$ \cite{hauke10,peter12}.
Another route to simulate quantum magnetism of dipolar systems, using the rotational structure of the ultracold polar molecules, has been
proposed in Refs. \cite{Barnett06, Gorshkov11}. In this case, the XY (or tunneling) term is restricted to nearest neighbors (NN) and only the ZZ 
(or interaction) term is dipolar.

Precisely this long-range (LR) dipolar character makes this system
interesting and challenging. Dipolar interactions introduce new
physics to conventional short-range (SR) systems. For example, for 
soft-core bosons, the system with long-range interactions and short-range
tunneling, can -- apart from Mott-insulating (MI), superfluid
(SF), and crystal phases -- host a Haldane-insulating phase
\cite{altman06}. This is characterized by antiferromagnetic (AFM)
order between empty sites and sites with double occupancy, with an
arbitrarily long string of sites with unit occupancy in
between\footnote{This is in contrast to standard insulating
phases, where the length of the string is fixed by the filling
factor.}. Another case where dipolar interactions play a crucial role is the appearance of the celebrated supersolid. The extended
Bose-Hubbard model (i.e., with NN interactions) with soft-core on-site interactions shows stable
supersolidity in one-dimensional (1D) \cite{Batrouni06}
and square lattices \cite{Sengupta05}, where in the 1D case a
continuous transition to the supersolid phase occurs, in contrast
to the first-order phase transition in two-dimensional (2D)
lattices. Long-ranged dipolar interactions on the other hand allow
for the appearance of supersolidity even in the hard-core limit
\cite{capogrosso10,pollet10}. In this limit, dipolar interactions
give rise to a large number of metastable states \cite{menotti07,
trefzger08} (for a review see \cite{lahaye09}). By tuning the
direction of the dipoles, incompressible regions like devil's
staircase structures have been predicted in Ref.~\cite{Ohgoe12}. While
being interesting, long-range interactions make computer
simulations of such systems quite difficult. On the other hand,
since ions in optical lattices may be extremely well controlled,
they form an ideal medium for a quantum simulator.

In our previous paper \cite{hauke10}, we considered both the
mean-field phase diagram for the system (\ref{hamspin}) as well as
a 1D chain using different quasi-exact techniques, such as density
matrix renormalization group (DMRG), and exact diagonalization for
small systems. Here, we would like to concentrate on frustration
effects on a 2D triangular lattice using a quantum Monte Carlo
(QMC) approach.

For the system consisting of spin-half particles (as, e.g.,
originating from two internal ion states in the Porras--Cirac
\cite{porras06} proposal)  the spins can be mapped onto a system of
hard-core bosons, using the Holstein--Primakoff transformation,
$S_{i}^{z} \rightarrow n_{i}-1/2$, $S_{i}^{+} \rightarrow
a_{i}^{\dag}$, and $S_{i}^{-} \rightarrow a_{i}$,
where $a_{i}^{\dag}$ $(a_{i})$ are the creation (annihilation)
operators of hard-core bosons and $n_{i}$ is the number operator at
site $i$.  These bosons obey the normal bosonic commutation 
relationships, $[a,a^{\dag}]=1$, but are constrained to only single filling, $a^{2}=(a^{\dag})^{2}=0$. 
This is the limit when the on-site repulsion term $U$ goes to infinity in the standard Bose-Hubbard model. 
In this representation, a spin up particle is represented
by a filled site and a spin down particle by an empty
site.  For convenience, we will mostly use the language of hard-core
bosons in this paper.

Before presenting the results for the 2D triangular lattice, let us briefly review
the behavior of the system for a 1D chain of bosons as obtained in Ref.~\cite{hauke10}.

\section{Review of the results in one dimension}

\begin{figure}
\centering
\includegraphics[width=\columnwidth]{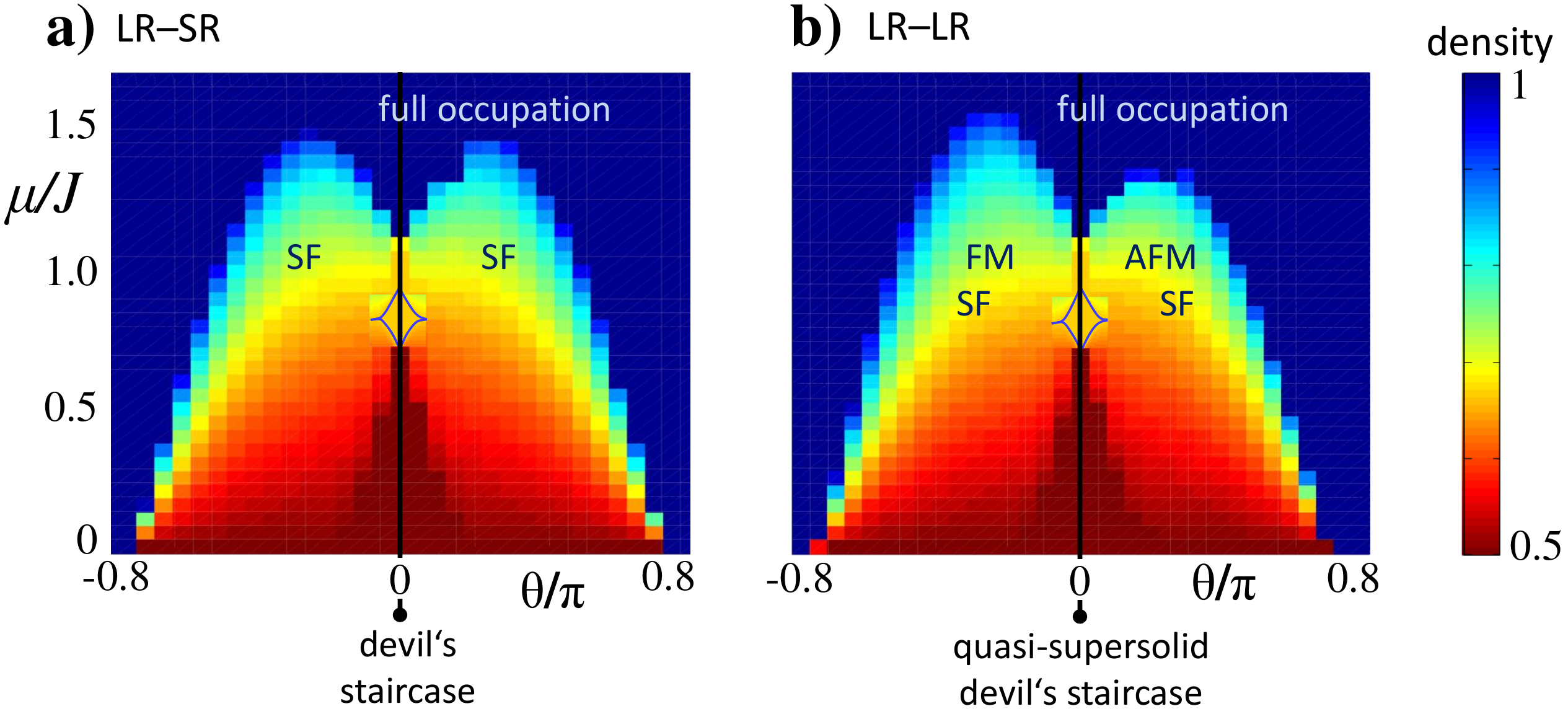}
\caption{
    \label{fig:PhaseDiagram1D}
    Phase diagram of the 1D system with dipolar interactions and
    {\bf (a)} nearest-neighbor tunneling,
    {\bf (b)} dipolar tunneling.
    The phases are labeled according to density matrix renormalization results, while the actual data shown comes from the infinite
time evolving block decimation method (with interactions truncated at next-to-nearest-neighbors), both from Ref.~\cite{hauke10}.
Along the black line, there exists a devil's staircase of crystal states. At finite tunneling, for (a), these spread into conventional
insulating states, while for (b), they form quasi-supersolids. The light blue lines near the center of the plots 
sketch the crystal lobes at $2/3$-filling. Their
cusp-like structure is typical for 1D systems. In 2D, they are expected to be rounded off, similar to Mott-lobes of the Bose--Hubbard model \cite{Freericks96}.
For nearest-neighbor tunneling, (a), the superfluid (SF) phases can be mapped into one another, while for nearest-neighbor dipolar
tunneling, (b), they are distinct on the ferromagnetic (FM, $\theta<0$) and the antiferromagnetic side (AFM, $\theta>0$).
Note also that in (b) frustration leads to an asymmetry between $\theta<0$ and $\theta>0$.
    }
\end{figure}

In order to form some intuition about possible physical effects
due to the LR interaction and tunneling, we now discuss briefly
the ground-state phase diagram of the Hamiltonian in 1D. For 
hard-core bosons, which is the topic of this article, the one-dimensional version of the Hamiltonian has been thoroughly 
investigated in the past \cite{hauke10} (see also \cite{Bak82,Deng05} 
for the special cases $\theta=0$ and $\theta=\pi/2$). For reference, 
we reproduce in Fig.~\ref{fig:PhaseDiagram1D} the phase diagrams for 
the system with dipolar interaction and NN tunneling, as well as the 
system with both the interaction and the tunneling terms dipolar.

At zero tunneling, the ground states are periodic crystals where -- to minimize the dipolar interaction energy -- occupied
sites are as far apart as possible for a given filling factor \cite{Bak82}. For finite 1D systems and very small tunneling such a
situation persists as exemplified in \cite{maik11}. For infinite chains in 1D, every fractional filling factor $n=p/q$ is a stable
ground state for a portion of $\mu$ parameter space. The extent in $\mu$ decreases with $q$, since at large distances the dipolar
repulsion is weak  and thus cannot efficiently stabilize crystals with a large period. This succession of crystal states is termed
the devil's staircase. This name derives from its surprising mathematical properties, challenging naive intuitions about continuity and measure:
since all rational fillings are present, it is a continuous function; moreover, its derivative vanishes almost everywhere
(i.e., it is non-zero only on a set of measure zero) -- and still it is not a constant, but covers a finite range.

At finite tunneling, the crystals spread into lobes similar to the Mott lobes of the Bose--Hubbard model. If the tunneling is only
over NNs, these Mott lobes are not sensitive to the sign of the tunneling and they form standard insulating states with diagonal
long-range order (LRO) and off-diagonal short-range order (Fig.~\ref{fig:PhaseDiagram1D}a). For long-range dipolar tunneling, the extent of the lobes is asymmetric
under sign change $\theta\to-\theta$: frustration effects stabilize the crystal states for $\theta<0$,  while the ferromagnetic
(FM) tunneling for $\theta>0$ destabilizes them (Fig.~\ref{fig:PhaseDiagram1D}b). Moreover, the crystal states acquire off-diagonal correlations which follow
the algebraic decay of the dipolar interactions \cite{hauke10}. This coexistence of diagonal and off-diagonal (quasi-)LRO turns
the crystal states into (quasi-)supersolids. The occurrence of such phases for hard-core bosons in 1D is truly exceptional, since
the systems where it appears consist typically  of soft-core bosons \cite{Batrouni06,Sengupta05}, or two-dimensional lattices \cite{capogrosso10}.
Furthermore, this 1D quasi-supersolid defies Luttinger-liquid theory, which typically describes 1D systems very well, even in the
presence of dipolar interactions \cite{Citro07,Citro08,Dalmonte10}. Where Luttinger-liquid theory applies, diagonal and off-diagonal
correlations decay algebraically with exponents which are the inverse of one another. Therefore, if the diagonal correlations show LRO,
the corresponding exponent is effectively $0$, and the exponent for the off-diagonal correlations is infinite, describing an exponential
decay. In our case, this exponent remains finite in the quasi-supersolid phase and the above relationship clearly does not hold.

At even stronger tunneling strengths, the crystal melts and the system is in a SF phase. The LR tunneling and interactions influence the
correlation functions at large distances and therefore also modify the character of this phase \cite{hauke10,Deng05}.

These results show that in this system the dipolar interactions considerably modify the quantum-mechanical phase diagram.
In higher dimensions, we can expect the influence of long-range interactions to be even stronger, which makes extending these studies
to a two-dimensional lattice highly relevant. For example, one can expect that -- if quasi-supersolids appear already in 1D -- the
long-range tunneling has a profound effect on the stability of two-dimensional supersolids, which appear in triangular lattices
at the transition between crystal and superfluid phases \cite{Batrouni06,Wessel05b,Heidarian05,Melko05}. Also, the frustration effects
already observed in the 1D system should be much more pronounced in the triangular lattice, simply due to the increased
number of interactions\footnote{In fact, we find that for positive hopping, $\theta>0$, our method of choice, quantum Monte Carlo,
fails due to the sign problem, invoked by frustration. See Section~\ref{cha:results} for details.}.

Further, such an analysis is especially relevant at finite temperature. In fact, a recent work scanned the phase diagram of Hamiltonian
\eref{hamspin} along the line $\mu=0$ in a square lattice \cite{peter12}. There, the authors found that above the superfluid on the FM
side (i.e., $\theta<0$), the continuous U(1) symmetry of the off-diagonal correlations remains broken even at finite temperatures.
Thus, the long-range nature of the tunneling leads to a phase which defies the Mermin--Wagner theorem \cite{Mermin66}\footnote{The
theorem remains valid, of course, as it applies only to short-range models.}.

For these reasons, we can expect intriguing effects of the long-range tunneling for the two dimensional triangular lattice, to which we turn now.

 \section{Dipolar hard-core bosons on the triangular lattice}

To analyze the system of dipolar bosons on the two dimensional
triangular lattice, we employ quantum Monte Carlo (QMC)
simulations at a finite but low temperature. Specifically, we study
rhombic lattices with periodic boundary conditions and $N=L \times
L$ sites, with $L=6..12$.  We will in particular thoroughly
investigate the Wigner crystal at $2/3$-filling. The triangular
lattice is frustrated even with only NN interactions.  When any
longer ranged interactions are added, this only increases the
frustration, which also makes QMC calculations more difficult.
Also, in frustrated systems \cite{Lhuillier05} and systems with
long-range interactions \cite{trefzger08} typically many
metastable states appear, making finding the ground state a
somewhat difficult task. To avoid complications with the sign
problem, caused by negative probabilities in QMC codes, we will
take only negative values of $\theta$ under consideration. Note
that this sign problem appears only for the long-ranged XY
interactions, while the ZZ (Ising-like) interactions are --
despite frustration -- sign-problem free.

In this study, we compare the Hamiltonian with both long-range
interactions and hoppings (LR--LR), with long-ranged dipolar
interactions but NN hopping (LR--SR; relevant for polar molecules
\cite{Burnell09b}), as well as with hopping and interactions
truncated at NNs (SR--SR; this is the NN XXZ model, relevant to
magnetic materials with planar anisotropy in their
couplings)\footnote{For all long-range terms, we truncate the
interactions at distances where they first reach the boundary of
the rhombic lattice. For the smallest system with $L=6$, this
amounts to including interactions up to distances of the
fifth-nearest neighbor.}.
Each of these systems will display different crystal, superfluid,
and supersolid regions. Comparing these cases will give valuable
insight into the influence of the long-range terms.

All the calculations were performed using the worm algorithm of
the open source ALPS (Algorithms and Libraries for Physics
Simulations) project \cite{ALPS}. This algorithm, first created by
N. Prokof'ev, works by sampling world lines in the path integral
representation of the partition function in the grand canonical
ensemble \cite{prokofev98}.

\subsection{Vanishing tunneling\label{cha:results}}

\begin{figure}

  \centering
  \includegraphics[width=1.0\textwidth]{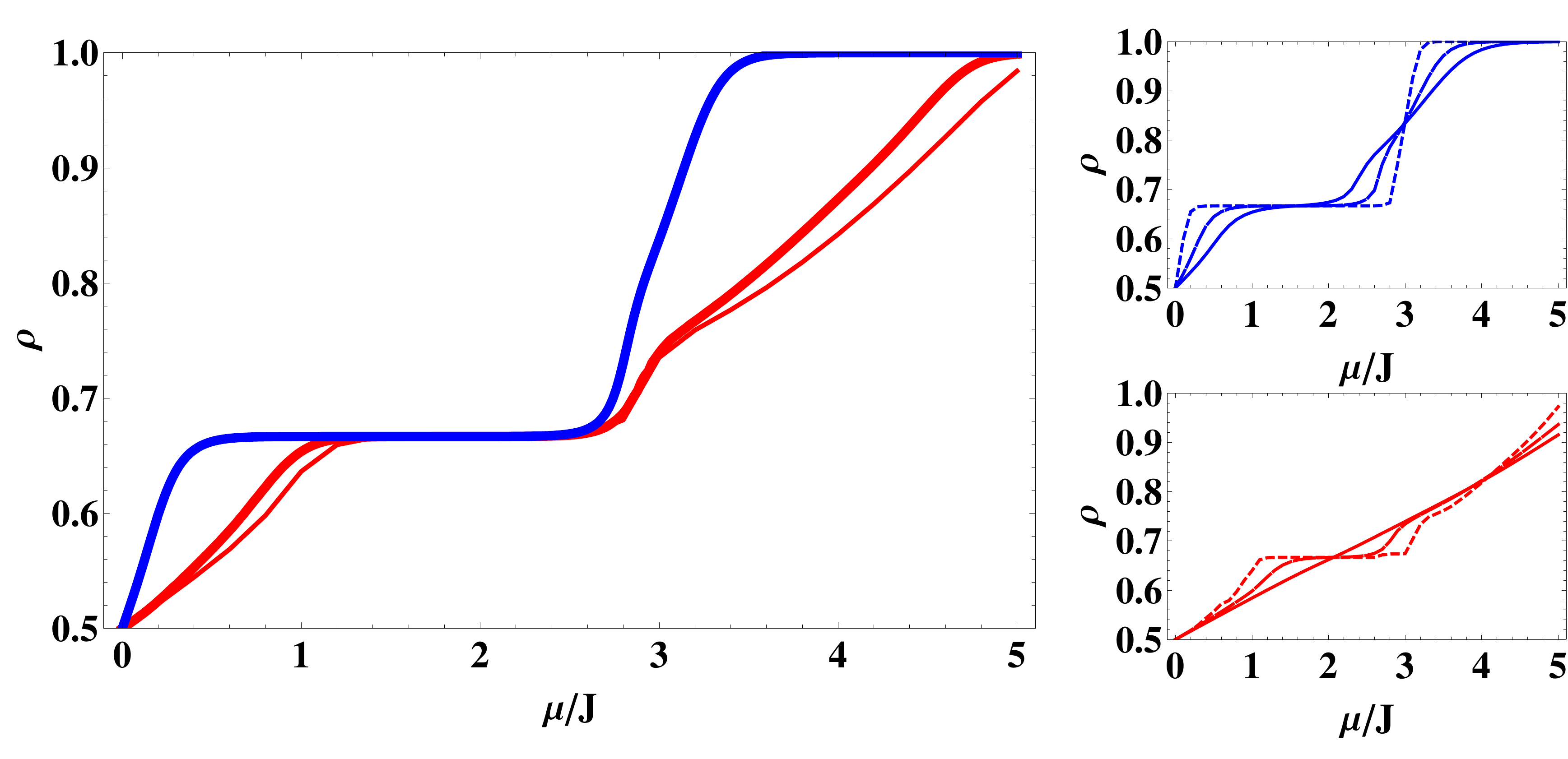}
  \caption{The graph on the left displays $\theta=0$ and $T=0.1$, where the density shows a single plateau for $2/3$-filling.
	For SR interactions (solid blue line), different curves for $L=6,9,12$coincide, and for LR interactions (solid red: 
	$L=6$, dashed red: $L=9$ and $12$) the size dependence is small. The panels 
	on the right are at fixed $L=12$ and $\theta=0$ for different temperatures $T=0.05, 0.15, 0.25$ (from dotted to dashed to solid). 
Both for SR (top right) and LR interactions (bottom right), the $2/3$-filling plateau shrinks as T increases.
}
  \label{devil}

  \end{figure}

The first calculation, which creates the motivation for the rest
of the paper, is to look at the case of vanishing tunneling and
temperature for each system (LR--LR, LR--SR, and SR--SR). Here,
similar to the 1D devil's staircase, at vanishing temperature a
series of insulating crystal states is expected to cover the
entire range of $\mu/J$. Since we are interested in finite
temperature results, we set $T=0.1$ -- which should still be low
enough to reflect the characteristics of the ground-state phase
diagram -- and look for plateaus in the density.  We distinguish
short- and long-ranged ZZ interactions.  From Fig.\ \ref{devil}, left panel, we
can see that the only plateau (besides the completely filled
system) that appears is at $\rho=2/3$ (corresponding either to $2/3$ boson filling or in spin terms, a lattice
with $2/3$ of the spins oriented up and $1/3$ oriented down) for both short- and
long-ranged interactions. Scaling the system size from $L=6$ to $L=12$ causes no change for
the short-ranged interactions, and minimal change for the
long-ranged ones.  The key difference is in the size and position
of the short-ranged and the long-ranged plateaus.  For
short-ranged interactions this plateau is larger and centered
around $\mu/J \simeq 1.5$, while the long-ranged interactions have
a smaller plateau centered around $\mu/J \simeq 1.85$. The finite width of these plateaus suggests that a $2/3$-filling Wigner crystal persists also
for some finite $\theta$.  
The right panels of Fig.\ \ref{devil} show how the plateau shrinks with increasing temperature, as well for SR interactions (top right panel) as 
for LR interactions (bottom right panel). In the latter case, in fact, by $T=0.25$ the plateau
has completely disappeared. 
We can also notice that at $T=0.05$ 
there are signs of some of the other plateaus, most noticably the $3/4$-filling plateau. 
The rest of the paper will focus on the $2/3$-filling crystal lobes and their properties.

\subsection{Low-temperature phase diagram at finite tunneling}

We now introduce a finite tunneling by choosing $\theta<0$ in our Hamiltonian, and study the properties around the $2/3$-filling crystal.
We calculate the density and the superfluid fraction.
The superfluidity is measured using the winding numbers calculated
from the movement of the worms in the QMC code.  In order to get
this value the system must have periodic boundary conditions so
that the world lines can properly ``wind'' around the system.
The superfluid fraction is

\begin{eqnarray}
  \rho_{s}=\frac{\langle W^{2} \rangle }{4 \beta},
  \label{SF}
\end{eqnarray}

 \noindent where $W$ is the winding number fluctuation of the world lines and $\beta$ is the inverse temperature.

\begin{figure}[hb]
  \begin{center}

   \includegraphics[width=1.0\textwidth]{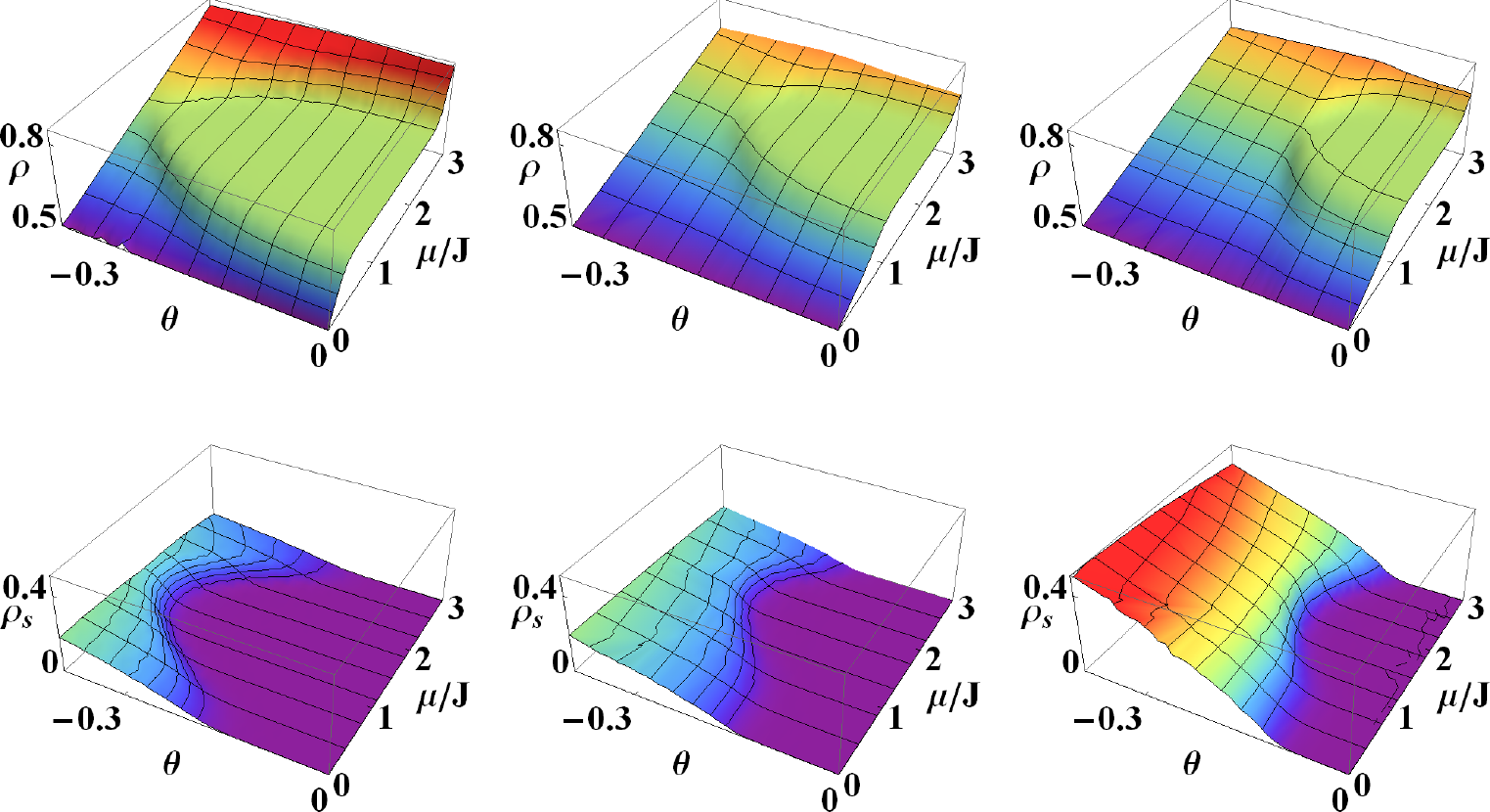}

  \caption{The columns show the $\rho=2/3$ lobes, evidenced by particle density (top row) and the superfluid fraction (bottom row) that arise under varying the ratio, $\theta$, and the chemical potential,
   $\mu/J$ (data for $L=6$).  The left column corresponds to the SR--SR system,  the middle one shows the LR--SR system and the right column depicts the lobes for the LR--LR system.
   Long-ranged dipolar interactions decrease the lobes in $\mu/J$ due to the appearance of devil's staircase like features, and long-range tunneling decreases the extent
   of the lobe in $\theta$.}
  \label{lobes}

  \end{center}
  \end{figure}

 Figure \ref{lobes} shows the results for the boson density of a $L=6$ triangular lattice at $T=0.1$.  For all types of interactions
we see that as $\theta$ increases in absolute value the $\rho = \frac{2}{3}$ plateau
shrinks, because larger $|\theta|$ increases the ratio of hopping to
dipolar interactions.  This introduces more kinetic energy and the
crystal melts into a superfluid.  It can be seen for the SR--SR
system that the boson density lobe extends to $\theta \simeq -0.36$,
while for the LR--SR interactions it ends at $\theta \simeq
-0.3$, and finally for LR--LR interactions the lobe is smaller
still, only going out to $\theta \simeq -0.2$.  The behavior is
explained by the fact that the increased amount of interactions
cause a quicker melting of the lobe.

 A second major observation is the shift in the position of the lobes. While the short-ranged lobe exists
approximately for $0.3 < \mu/J < 2.7$, the long-ranged lobes lie generally on the interval $1.0 < \mu/J < 2.8$.  For a system
like this at $T=0$, one expects the $\rho = \frac{2}{3}$ lobe to exist on the range $0 < \mu/J < 3$ with a mirrored image of the $\rho = \frac{1}{3}$ lobe
at $-3 < \mu/J < 0$.  The existence of the two identical lobes is
explained by particle-hole symmetry
\cite{Wessel05b,Heidarian05,Melko05}. The lobes are separated with
a kind of mixed solid in between (with coexistence of $1/3$- and
$2/3$-filling regions) The existence of this region may be caused
by several phenomena.  It could either be an effect of the finite
temperature as in Ref.\ \cite{Ozawa12}, or due to the
existence of many metastable states caused by a devil's
staircase like behavior, similar to what was observed in Ref.\ \cite{Ohgoe12}.

 Again referring to the $T=0$ phase diagram, we expect that there is a region of supersolidity that extends in between the lobes
and goes all the way to their base at $\theta=\mu=0$ \cite{Wessel05b}.  In our system this region should exist near the tip of the lobe but not extend all
the way to the base due to the finite temperature and the resulting mixed solid.  Looking at Fig.\ \ref{lobes}, it is obvious indeed that, if a
supersolid region exists, it can only be near the tip of the lobe because the superfluidity is zero a significant way up the lobe.  Judging by the increased
separation of the long-ranged lobes, we can assume that the supersolid region for these systems should increase in size to fill the region in between.
To search for the supersolid phase, we now compare the superfluidity with the static structure factor.

The structure factor is defined as the Fourier transform of the density-density correlations,

 \begin{eqnarray}
   S(\mathbf{Q})=\left\langle\left\vert \sum\limits_{i=1}^{N}n_{i}e^{\imath \mathbf{Qr}_{i}} \right\vert^{2}\right\rangle/N^{2}.
   \label{ST}
 \end{eqnarray}

Here, we focus on the wave vector ${\bf Q}=(4\pi/3,0)$, which corresponds to the $\sqrt{3} \times \sqrt{3}$ order parameter that is
associated with $1/3$- and $2/3$-filling crystals on the triangular lattice. 
For the case of the $2/3$-filling lobe that we are interested in, it will show plateaus over the same range of $\mu$ as the density, 
but additionally gives insight into the arrangement of the bosons on the lattice. 
This makes it a useful quantity in searching for supersolid regions. 
In fact, a supersolid exists when both the structure factor and the superfluid
fraction have non-zero values. The physical mechanism behind the supersolid phenomenon is based upon the appearance of extra holes
(particles). The underlying crystal structure has $\sqrt{3} \times \sqrt{3}$ order on a triangular sublattice of the physical lattice.  The extra
holes (particles) are free to move around on the rest of the lattice as superfluid objects. In this way, the system retains a crystal structure,
while it acquires at the same time the long-range coherence of a superfluid. Due to the hole (particle) doping, it forms in sections away
from commensurate filling, in this case in between the $1/3$- and $2/3$-filling lobes.

  \begin{figure}[ht!]
  \begin{center}

   \includegraphics[width=1.0\textwidth]{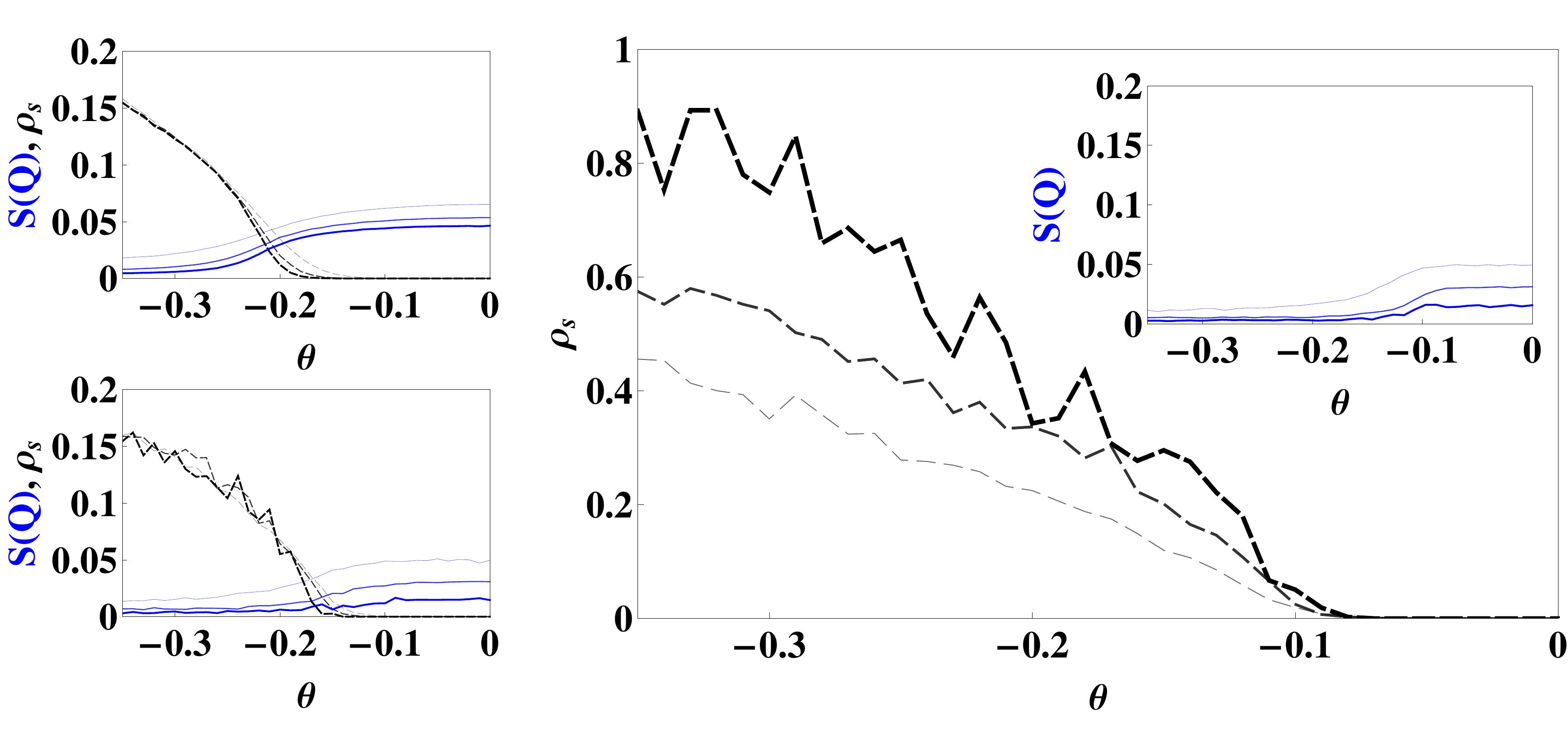}

  \caption{Cuts at $\mu/J=0$ for SR--SR (left top), LR--SR (left bottom) and LR--LR (right) for a $L=6, 9$, and $12$ lattice (lines become 
thicker and darker with increasing system size).  
For all cases, the structure factor (solid blue) is finite at small $\theta$ and the superfluid fraction (dashed black) at large $\theta$. 
At the system sizes studied, there appears a supersolid region at intermediate $\theta$ where both structure factor and superfluid 
fraction are finite. In the LR--LR system there is a reversal of finite size effects. In this case the superfluid fraction 
for larger systems becomes higher instead of lower.}
  \label{horizSS}

  \end{center}
  \end{figure}

 Taking ``slices'' out of the crystal lobes we now check where there is a supersolid region and where the system transitions
directly from crystal to superfluid.  Also we study the nature of these transitions to see if they are of first or second order.

 The most logical region to look for supersolids is directly in between the $1/3$- and $2/3$-filling lobes, at $\mu/J=0$.  Figure \ref{horizSS}
shows the behavior of the SR--SR, the LR--SR, and the LR--LR system at this cut for several system sizes.  Structure factor and superfluidity reveal, for
all the systems, three different regions.  In each case, the system starts at $\theta=0$ in a solid phase where the superfluid fraction is
zero but the structure factor is finite.  It transitions smoothly into a supersolid region where both superfluidity and structure
factor are non-zero.  Finally, the structure factor smoothly drops away and leaves just a non-zero superfluid fraction, making the final
phase a superfluid.  In each system, the size of the supersolid region is different. In the SR--SR case, the supersolid region begins to
appear at $\theta \simeq -0.15$ for a $L=6$ lattice.  As the size grows to $L=12$, the region has shifted to $\theta \simeq -0.19$ with the
superfluid curves becoming sharper.  The increased system size also reduces the value of the structure factor a little. From \cite{Heidarian05}, we know
that at even larger sizes (but $T=0$) the supersolid will continue to exist in this type of system.
For the LR--SR system the supersolid region appears at a similar point and also shifts with system-size increase.  The structure factor on the
other hand has a significant decrease for larger system sizes. It is difficult to tell if at greater sizes the existence of the supersolid will persist.
The final graph shows the LR--LR system.  In this system, superfluidity appears even before $\mu/J$ reaches $-0.1$.  In this system, the superfluid fraction
is much greater than in the previous two because of the long-ranged tunneling.  This means that at small system sizes the supersolid region is much more prominent
relative to the crystal lobe. The structure factor diminishes with system size almost exactly as in the LR-SR case except that the transition is at a different
value of $\theta$, and near $\mu/J=0$ it drops to slightly lower values.  Due to this strong decrease, for any situation with long-range interactions we cannot
clearly state whether the supersolid region survives at larger system sizes.

  \begin{figure}[ht!]
  \begin{center}

   \includegraphics[width=1.0\textwidth]{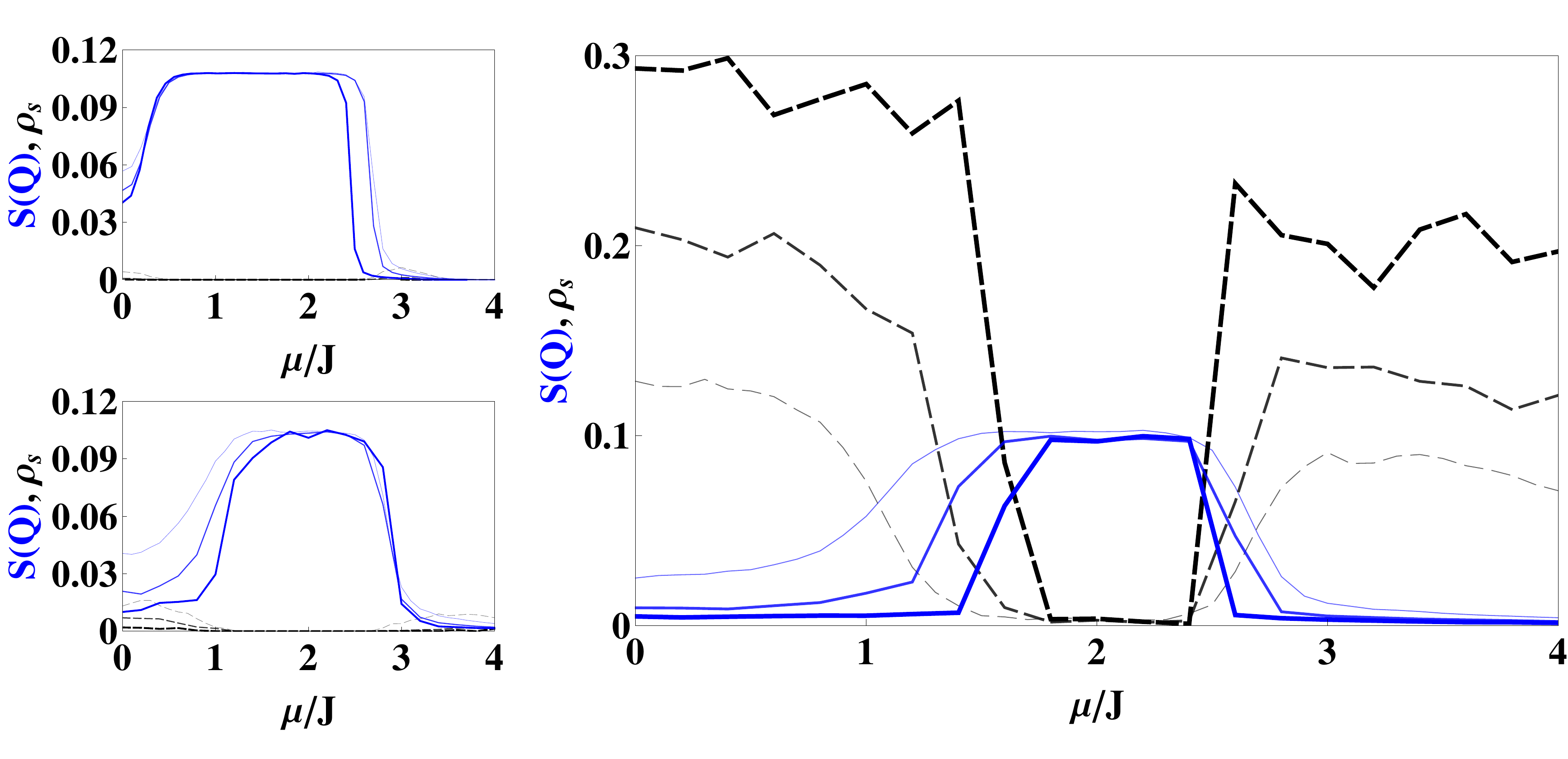}

  \caption{Cuts at $\theta=-0.15$ for SR--SR (left top), LR--SR (left bottom), LR--LR (right) for a $L=6, 9$, and $12$ lattice 
(lines become thicker and darker with increasing system size).  
Solid blue: Structure Factor.
Dashed black: Superfluid fraction.
At this value of $\theta$, for short-range tunneling, the superfluid fraction disappears rapidly with increasing system size, 
while for long-range tunneling it even increases. At $\mu/J\approx 1.75$, possibly a supersolid may survive in large lattices.
Again in the LR--LR system there is a reversal of finite size effects.  The superfluid fraction for larger systems 
becomes higher instead of lower. 
}
  \label{vertcut15}

  \end{center}
  \end{figure}

 Next we take vertical cuts at a value of $\theta=-0.15$, since this is a reasonable place for a supersolid to exist for the LR-LR system ($\simeq80\%$ of the
tip of the lobe). We compare all the systems at this cut, and study the behavior of the superfluid
fraction and the structure factor, plotted in Fig.\ \ref{vertcut15}.  For SR--SR interactions, no supersolid region appears.  On one side of
the lobe there is a sharp phase transition directly from the crystal to the superfluid phase, while on the other side there is a slower change
from one solid form to another ($\rho=1/3\to2/3$). The finite-size scaling in the figure shows that as the size increases the transitions of the
structure factor become even sharper, although they stay continuous due to the finite temperature.  The values of the superfluid fraction decrease
as the system sizes increases and essentially disappear at $L=12$.  In the LR--SR system, a hint of the supersolid
phase begins to appear on either side of the lobe.  It is a bit more evident on the side where $\mu/J$ is small (as is to be expected from references such as
\cite{Wessel05b}), but it also arises on the opposite side. This is contrary to a system of only short-ranged interactions where this supersolid region
appears only on one side of the lobe and not both. At larger sizes, also in the LR--SR system the transitions become sharper and the superfluidity gets smaller.
The final and most interesting cut is taken out of the LR--LR lobe.  In this system, we see a smooth transition from crystal to supersolid at $\mu/J \simeq 2.4$
and at $\mu/J \simeq 1.4$ for $L=6$. For larger systems the transition at $\mu/J \simeq 2.4$ occurs at the same spot but becomes sharper, making the
supersolid region disappear.  At $\mu/J \simeq 1.4$ the transition shifts to a higher value of $\mu/J$, making the $2/3$-filling plateau smaller. It also becomes
less smooth but the supersolid region remains longer than for the SR--SR case. In the LR--LR system,
the superfluid fraction for larger systems has the opposite effect than for the previous cases, it becomes higher instead of lower.

  \begin{figure}[ht!]
  \begin{center}

  \includegraphics[width=0.49\textwidth]{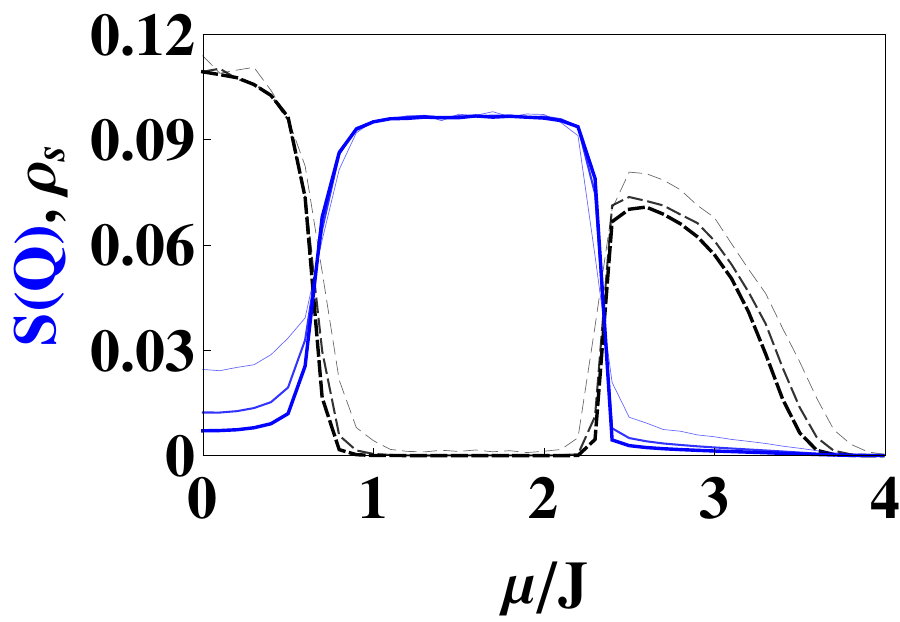}
  \includegraphics[width=0.49\textwidth]{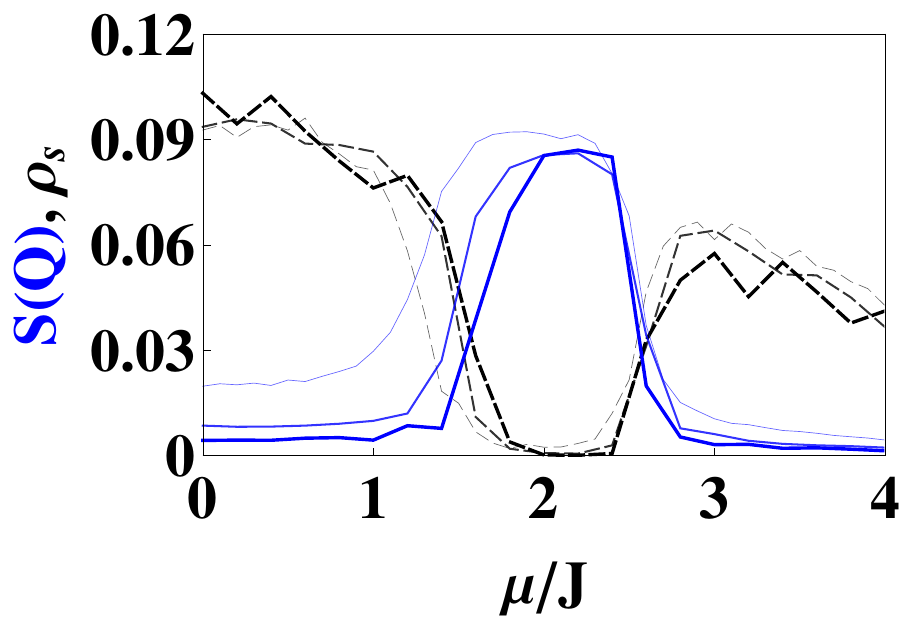}

  \caption{SR--SR (left) at $\theta=-0.28$ and LR--SR (right) at $\theta=-0.23$.  Solid blue: Structure Factor.
Dashed black: Superfluid fraction.  Lines become thicker and darker as system size goes up.
In the SR--SR system, a supersolid at small $\mu/J$ persists at large systems, while at $\mu/J\approx2.4$ the transition from crystal to superfluid becomes a direct first-order transition. For the LR--SR system, the curves change slowly at both sides of the crystal lobe, leading to persisting supersolids. 
}
  \label{vertSS}

  \end{center}
  \end{figure}

A perhaps fairer comparison is to look at a cut through a region where we are sure the supersolid exists for all three systems. Therefore, in
Fig.\ \ref{vertSS} we look at two more cuts that are now taken closer to the tips of the SR--SR and LR--SR lobes. As with the LR-LR system (last panel of
Fig.\ \ref{vertcut15}), these lie at around $80\%$ of the tip of the lobe. In the SR--SR lobe,
the cut is taken at $\theta =-0.28$.  Here we see a similar behavior for the structure factor as we did in the $\theta =-0.15$ cut of
the lobe, but this time the superfluid fraction plays a much more important role.  On the one side, $\mu/J \simeq 2.4$, both the superfluid
fraction as well as the structure factor have sharp transitions that become even sharper at larger sizes.  In fact, at $T=0$ these transitions
have been shown to be of first order, and the system goes directly from crystal to superfluid.  Due to the finite temperature, they are continuous in our case.  
On the other side, where $\mu/J \simeq 0.8$, there appears a second order phase transition into
a supersolid region that spans all the way to $\mu/J = 0$.  Finally, we take a cut at $\theta=-0.23$ of the LR--SR lobe.  The behavior
of this system seems to be quite different.  The first thing to notice is that the transitions on either side of the lobe are
of second order.  The other, and more important, observation is that now it appears that this system has supersolid behavior on both sides
of the lobe: in addition to the expected supersolid at smaller $\mu/J$, a region at $\mu/J$ above the crystal lobe appears where both structure factor and 
superfluid fraction are finite. If we recall Fig.\ \ref{vertcut15}, right panel, the LR--LR system showed that at $\theta =-0.15$ as $L$ increased the supersolid region
disappeared from the upper side of the lobe.  In the case of the LR--SR system for $\theta = -0.23$ the increase in the system size does not
get rid of this supersolid phase.

\subsection{Finite temperature results}

  \begin{figure}[hb!]
  \begin{center}

  \includegraphics[width=1.0\textwidth]{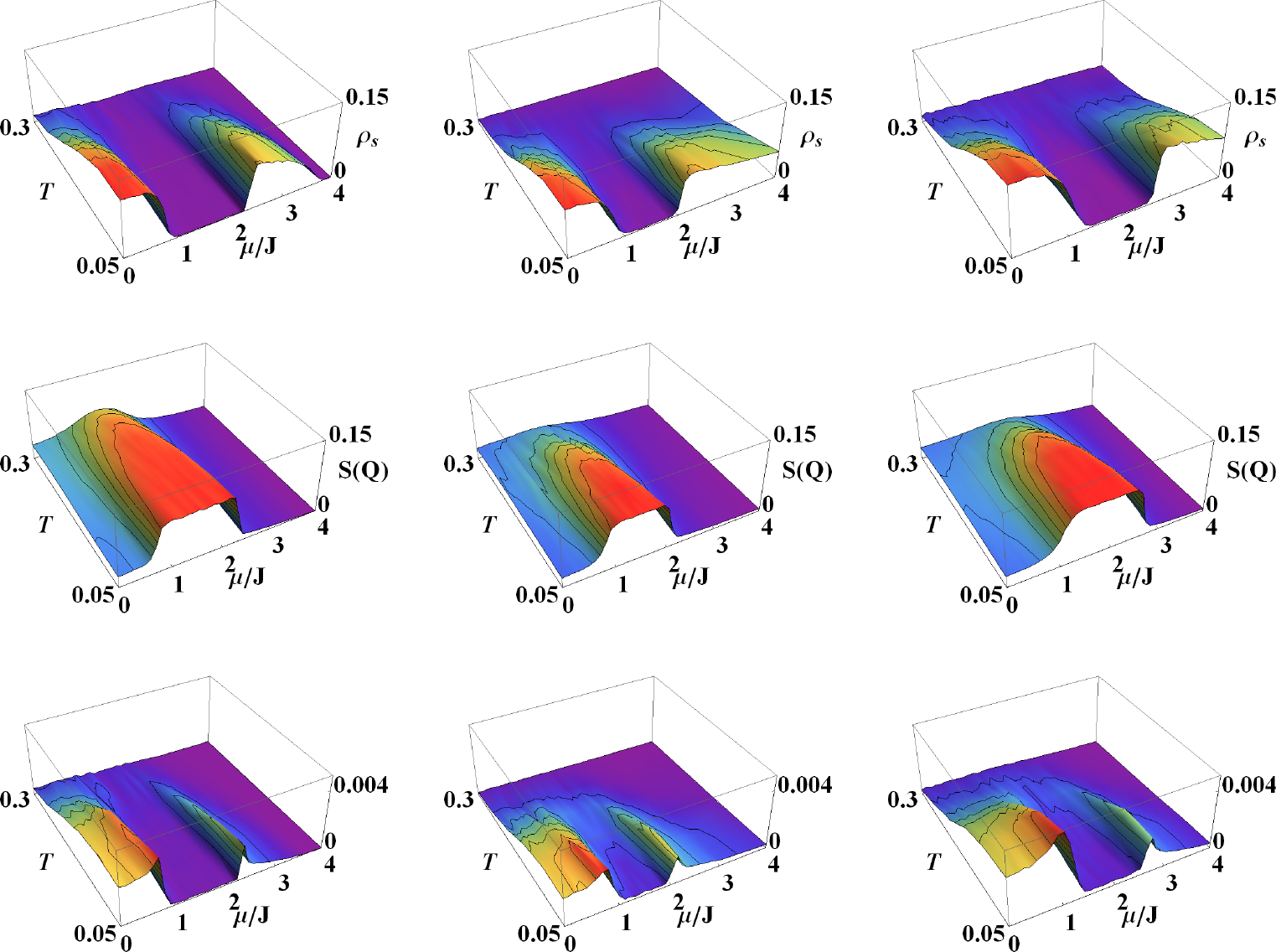}

  \caption{Each row shows a different object : the top row  -- the stiffness,  the middle row -- the structure factor while the bottom row represents a product 
 of the first two rows.  Columns  corresponds to different systems:  Left column is for
SR--SR at $\theta = -0.28$, middle column yields LR--SR at $\theta=-0.23$, right column is for LR--LR at $\theta=-0.15$.
For all three systems, the three distinct quantum phases -- crystal, supersolid, and superfluid -- survive over some temperature range before they melt. 
}
  \label{SSmelt}

   \end{center}
   \end{figure}

 As a final calculation, we take a look at the important role that the temperature plays in both the melting of the crystal
as well as the supersolid region.  In this section, we will use the same cuts as in the previous section ($\theta=-0.28$ for
SR--SR, $\theta=-0.23$ for LR--SR and $\theta=-0.15$ for LR-LR) so that each system will posses all the possible phases:
crystal, superfluid, and supersolid.  Each cut is investigated for $0.05 < T < 0.3$ at a system size of $L=6$.  First, 
we analyze the structure factor to study how the $\rho=2/3$ crystal melts with an increase in temperature (second row of Fig.\ \ref{SSmelt}).  For the SR--SR interactions, 
even at a temperature of $0.3$ there still exists a bump in the structure factor which indicates that the crystal has not completely
melted yet, while for both of the long-ranged lobes the crystal melts by $T \simeq 0.3$.  Interestingly, the SR--SR crystal and the LR--LR crystal are approximately the same size at $T=0.05$, but by $T=0.3$ one has melted while the other
still exists.  That means that the system with short-ranged interactions holds its crystal structure better at higher temperatures
than does our system with all long-ranged interactions.  Looking at the LR--SR lobe, we see that its crystal at this cut starts off
smaller, yet it melts at about the same temperature as the one for LR--LR interactions.  It seems that the dipolar repulsion helps
stabilize the crystal structure over a larger temperature range, while the long-ranged hopping destroys the crystal more quickly
because of the extra kinetic energy.

Maybe more importantly, we now study the melting of the supersolid for these same cuts.  Figure \ref{SSmelt} shows the
structure factor, superfluidity, and supersolidity as a function of temperature for each of the different systems.  Since the
supersolid is defined by having both non-zero structure factor and non-zero superfluid fraction, by combining the graphs
we are able to see where these regions exist and also how they melt with increased temperature (the bottom row of Fig.\ \ref{SSmelt} shows a product of structure factor and superfluid fraction, which remains finite only where the two coexist).
A common feature of all the graphs are the spikes on either side of the plateaus.  These are regions where a phase transition occurs
but does not necessarily imply that a supersolid region exists.  Most likely, these features appear due to the finite size of the
system and the resulting smooth transitions of superfluid fraction and structure factor.  At larger sizes, the transitions would be much sharper at these points, the regions where a finite structure factor and superfluid fraction coexist would shrink, and the spikes would diminish.  From the previous
section, we can assume that for the SR--SR system they would disappear completely at the upper transition from the crystal lobe while for the other two systems there would still exist a small supersolid region.

Returning to the main focus, the small-$\mu$ region, we see that in each case a supersolid region appears that extends from the left side of the
plateau all the way to $\mu/J=0$.  In every system, this supersolid region also exists for a finite range of temperatures.  For SR--SR
interactions, it gradually decreases but still extends all the way out past $T=0.3$.  For the LR--SR interactions, the
supersolid region again slowly melts but now disappears at $T \simeq 0.23$, just below the spot where the crystal melted.  The
supersolid region for the LR--LR system appears to have the largest magnitude of the three systems, but then rapidly melts at $T \simeq 0.3$.

  \begin{figure}[ht!]
  \begin{center}

  \includegraphics[width=0.32\textwidth]{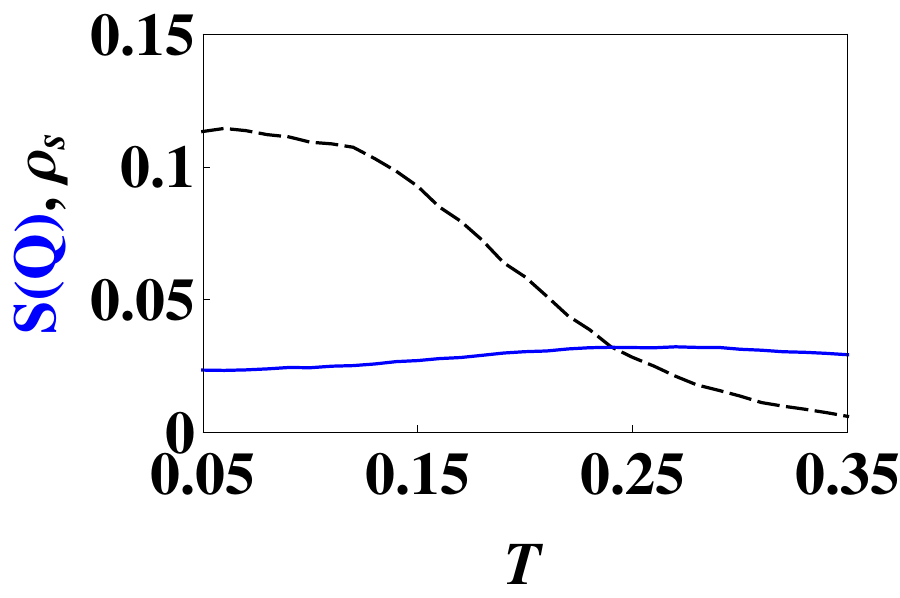}
  \includegraphics[width=0.32\textwidth]{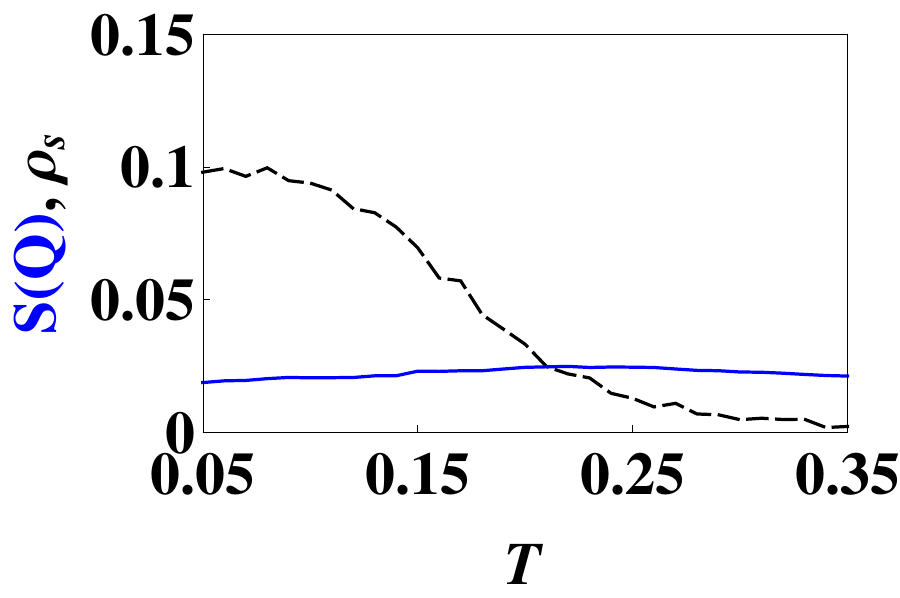}
  \includegraphics[width=0.32\textwidth]{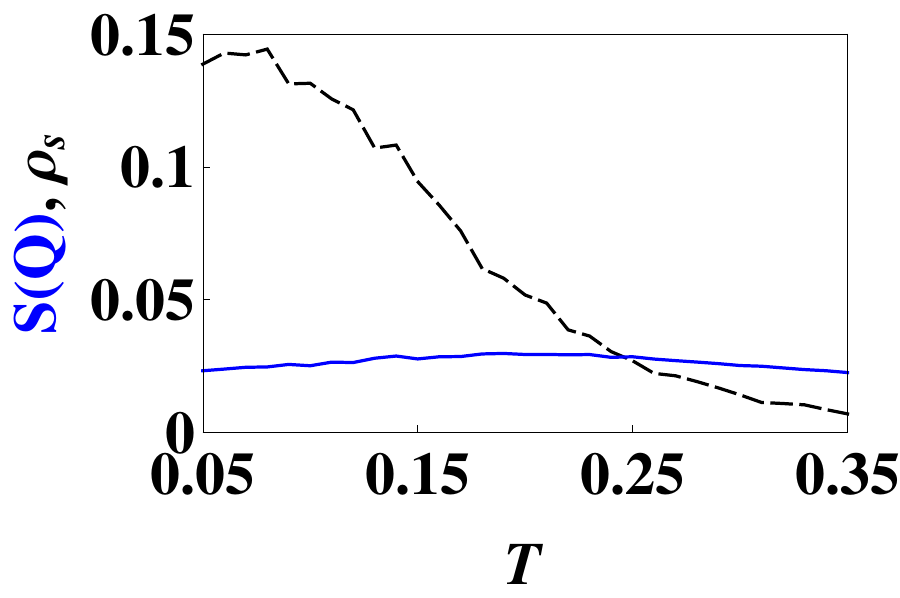}

  \caption{SR--SR at $\theta=-0.28$, LR--SR at $\theta = -0.23$ and LR--LR at $\theta=-0.15$ (left to right).  All cuts are taken at $\mu/J=0$. 
  The structure factor (solid blue) attains similar values for all three systems. The superfluid fraction (dashed black) is largest in the LR--LR 
  system and melts fastest in the LR--SR one.}
  \label{MUmelt}

  \end{center}
  \end{figure}

 In order to compare these transitions more quantitatively, we take a cut along $\mu/J=0$ for each system,
shown in Fig.\ \ref{MUmelt}. All three systems show a relatively similar and steady value for the structure factor.  Hence, the
values of the superfluid fraction are going to determine the existence of the supersolid regions.  The first
plot shows the SR--SR system at the $\theta=-0.28$ cut, and we can see that the superfluid fraction stays non-zero all the way
out to $T=0.35$.  The LR--SR system has a very similar behavior at the $\theta=-0.23$ cut, but in this case the supersolid is nearly
completely melted by $T=0.35$.  The final plot is the LR--LR system at $\theta=-0.15$, which behaves slightly differently.  The most
important difference is that the starting value of the superfluid fraction is higher than in the first two plots.  This should
therefore make the supersolid region more pronounced.  But even though the superfluid fraction has the highest value for this
system, it decays more quickly and reaches values similar to the SR--SR system at $T\approx 0.35$.

\section{Conclusion}

In this paper, we have presented a quantum Monte Carlo study of
dipolar spin models that describe various systems of ultracold
atoms, molecules, and ions. We have presented predictions
concerning the phase diagram of the considered systems at zero and
finite temperatures, and described the appearance and some properties
of the superfluid, supersolid, and crystalline phases. While the
results are not surprising and resemble earlier obtained results
for similar systems in 1D and 2D, the main advantage of our study
is that it is directly relevant to the current experiments:
\begin{itemize}

\item The results for the XXZ model with short-range tunneling
apply for ultracold gases of polar bosonic molecules in the limit
of hard-core bosons. Note that the earlier works
\cite{capogrosso10,pollet10} have concentrated on the appearance of
the supersolid phase and devil's staircase of crystalline phases
in the square lattice \cite{capogrosso10}, and supersolid and
emulsion phases in the triangular lattice \cite{pollet10}. Here we
focus on the hard-core-spin limit, and compare it and stress
differences with other models, such as the ones with long-range
tunneling, i.e., long-range XX interactions.

\item The results for the XXZ  models with long-range tunneling
apply for systems of trapped ions in triangular lattices of
microtraps. These results are novel, since so far
such models have been only studied using various techniques in 1D,
and using the mean-field approach in 2D. While the first
experimental demonstrations of such models were restricted to a few
ions (see for instance \cite{friedenauer08}),  many experimental
groups are working on an extension of such ionic quantum simulators
to systems of many ions in microtraps  \cite{priv}. In fact, very
recently the NIST group has engineered 2D Ising interactions in a
trapped-ion quantum simulator with hundreds of spins
\cite{britton12}. Although in this experiment the quantum regime
has not yet been achieved, it clearly opens the way toward quantum
simulators of spin models with long-range interactions. We expect that
in the near future the result of our present theoretical study
will become directly relevant for experiments.

\end{itemize}

 \section*{Acknowledgements}

This work was supported by the International PhD Projects
Programme of the Foundation for Polish Science within the European
Regional Development Fund of the European Union, agreement no.
MPD/2009/6. We acknowledge financial support from Spanish
Government Grants TOQATA (FIS2008-01236) and Consolider Ingenio
QOIT (CDS2006-0019), EU IP AQUTE, EU STREP NAMEQUAM, ERC Advanced
Grant QUAGATUA, CatalunyaCaixa, Alexander von Humboldt Foundation and Hamburg
Theory Award. M.M. and J.Z. thank Lluis Torner, Susana Horv\'ath and all ICFO personnel for
hospitality. J.Z. acknowledges support from Polish
National Center for Science project No. DEC-2012/04/A/ST2/00088.

\end{document}